# A uniformly moving and rotating polarizable particle in thermal radiation field: frictional force and torque, radiation and heating

G. V. Dedkov[1] and A.A. Kyasov

Nanoscale Physics Group, Kabardino-Balkarian State University, Nalchik, 360000, Russia

**Abstract.** We study the fluctuation-electromagnetic interaction and dynamics of a small spinning polarizable particle moving with a relativistic velocity in a vacuum background of arbitrary temperature. Using the standard formalism of the fluctuation electromagnetic theory, a complete set of equations describing the decelerating tangential force, the components of the torque and the intensity of nonthermal and thermal radiation is obtained along with equations describing the dynamics of translational and rotational motion, and the kinetics of heating. An interplay between various parameters is discussed. Numerical estimations for conducting particles were carried out using MATHCAD code. In the case of zero temperature of a particle and background radiation, the intensity of radiation is independent of the linear velocity, the angular velocity orientation and the linear velocity value are independent of time. In the case of a finite background radiation temperature, the angular velocity vector tends to be oriented perpendicularly to the linear velocity vector. The particle temperature relaxes to a quasistationary value depending on the background radiation temperature, the linear and angular velocities, whereas the intensity of radiation depends on the background radiation temperature, the angular and linear velocities. The time of thermal relaxation is much less than the time of angular deceleration, while the latter time is much less than the time of linear deceleration.

## 1. Introduction

Quantum zero-point fluctuations of electromagnetic field in vacuum lead to the Casimir attraction between neutral conducting bodies [1]. When the bodies are moving, they may pull off real photons from fluctuating QED vacuum [2]. In this relation, it is worthwhile to mention the phenomenon of Zel'dovich' superrradiance [3], when rotating cylinder amplifies certain waves of incident electromagnetic radiation. The interaction of small moving particles with thermal radiation was first considered by Einstein [4]. In dynamically and thermally nonequilibrium configurations related with objects in relative motion, the interacting bodies experience not only the action of attractive/repulsive forces and heating/cooling effects, but also frictional forces and mechanical torques [5-18] (it is worth noting that the reference list is far from completeness). Despite its minuteness, the exchange of momentum by light-matter interaction is responsible for various phenomena ranging from the formation of cometary tails and thermal radiation of cosmic dust to trapping and cooling of nanoparticles down to single atoms.

---
[1] Corresponding author e-mail: gv_dedkov@mail.ru

Unlike longstanding issues related with fluctuation-electromagnetic forces and heat exchange, investigation of thermal and nonthermal radiation effects produced by moving/rotating polarizable nanoparticles has a rather short history [19–30]. Thus, radiation produced by rotating particle in vacuum was studied in [19, 20, 22, 24, 29], the effect of uniform relativistic motion of small and big particles on the intensity of thermal radiation was investigated in [23, 27] and [30], whereas the quantum Cherenkov radiation at non-contact friction between the bodies in relative motion was studied in [22, 26, 28].

In the case of rotational-translational motion of particle in vacuum, the picture of radiation becomes more intriguing due to an interference between various physical factors such as linear and angular velocities, temperature of particle and background radiation, relative orientation of the vectors of linear and angular velocity. In this work, in contrast to [24], we treat the general case of relativistic translational-rotational motion of a small particle in vacuum background assuming arbitrary mutual orientation of the linear and angular velocity vectors. Within the framework of fluctuation-electromagnetic theory, we obtain general expressions for the dissipative tangential force acting on a particle, components of torque, the rate of heating and intensities of thermal and nonthermal radiation. The interrelation between different physical factors and their temporal dependences are discussed. It is shown that mutual orientation of angular and linear velocity vectors and the magnitude of linear velocity are independent of time at zero temperature conditions. However, the particle is heated since only a part of the kinetic energy of rotation is radiated. At finite temperature of particle and (or) vacuum background, the nonthermal radiation is generated along with thermal radiation. Moreover, the spin of particle tends to rotate perpendicularly to the direction of linear velocity. The linear deceleration occurs much slower than the rotational deceleration, and the times of deceleration turn out to be many orders of magnitude less than the time needed to reach the state of thermal quasiequilibrium. Some numerical estimations corresponding to conductive nanoparticles like graphite are presented. The system of Gauss units is used throughout the article.

## 2. Basic assumptions and theoretical relations

We accept quasistationary conditions assuming that a small neutral spherical particle with radius $R$ and local temperature $T_1$ (in the own frame of reference $\Sigma''$) moves with linear velocity $\mathbf{V}$ with respect to the reference frame $\Sigma$ of a vacuum background along the $x$ axis and rotates with angular velocity $\mathbf{\Omega} = \Omega\mathbf{n}$ in co-moving reference frame $\Sigma'$ (Fig. 1). The frame $\Sigma'$ also moves with velocity $\mathbf{V}$ along the $x$ axis in $\Sigma'$. Due to obvious asimuthal symmetry of all physical quantities relative to the direction of $\mathbf{V}$, the axes $(x', y', z')$ of $\Sigma'$ can be chosen so that the angular velocity vector $\Omega\mathbf{n}$ lies in the plane $(x', z')$.

A vacuum background is assumed to be filled with equilibrium photonic gas with temperature $T_2$. Moreover, during emission of low-frequency photons, we may treat the particle as being a dipole with fluctuating dipole and magnetic moments $\mathbf{d}(t), \mathbf{m}(t)$. This requires the conditions $\Omega R/c \ll 1$, $R \ll \min(2\pi\hbar c/k_B T_1, \ 2\pi\hbar c/k_B T_2)$, where $\hbar, k_B, c$ – are the Planck and Boltzmann constants and the speed of light in vacuum. Material properties of particle are described by the frequency-dependent dielectric/magnetic polarizabilities $\alpha_e(\omega)$, $\alpha_m(\omega)$ which are given in its own reference frame $\Sigma''$. Throughout the paper we use the Gauss units.

Let the surface $\sigma$ encircles the particle at a large enough distance so that the fluctuating electromagnetic field on $\sigma$ can be considered as the wave field. According to the energy conservation law for the system in the volume $\Omega$ (not to be confused with angular velocity $\Omega$) restricted by $\sigma$, one may write

$$-\frac{dW}{dt} = \oint_\sigma \mathbf{S} \cdot d\vec{\sigma} + \int_\Omega \langle \mathbf{j} \cdot \mathbf{E} \rangle d^3r \tag{1}$$

where $W = (1/8\pi)\int_\Omega \left(\langle \mathbf{E}^2 \rangle + \langle \mathbf{H}^2 \rangle\right) d^3r$ is the energy of fluctuating field in the volume $\Omega$, $\mathbf{S} = (c/4\pi)\langle \mathbf{E} \times \mathbf{H} \rangle$ is the Pointing vector, $\mathbf{E}$ and $\mathbf{H}$ are the corresponding electric and magnetic fields, and $\mathbf{j}$ the density current. The angular brackets denote the total quantum and statistical averaging and all physical quantities correspond to the reference frame $\Sigma$, if not specified.

In quasistationary regime ($dW/dt = 0$) one obtains from (1)

$$I = \oint_\sigma \mathbf{S} \cdot d\vec{\sigma} = -\int_\Omega \langle \mathbf{j} \cdot \mathbf{E} \rangle d^3r \equiv I_1 - I_2 \tag{2}$$

where $I_1 = I_1(T_1)$ is the intensity of radiation in vacuum, and $I_2 = I_2(T_2)$ –the intensity of radiation coming from vacuum and absorbed in the particle.

Using relativistic transformations of density current and charge, the electric field and volume in frames $\Sigma$ and $\Sigma'$, one obtains ($\beta = V/c$) [9]

$$\int_{\Omega'} \langle \mathbf{j}' \cdot \mathbf{E}' \rangle d^3 r' = \frac{1}{1-\beta^2} \left( \int_{\Omega} \langle \mathbf{j} \cdot \mathbf{E} \rangle d^3 r - F_x \cdot V \right) \equiv \frac{1}{1-\beta^2} dQ/dt, \tag{3}$$

$$dQ/dt = \langle \dot{\mathbf{d}} \cdot \mathbf{E} + \dot{\mathbf{m}} \cdot \mathbf{H} \rangle, \tag{4}$$

$$F_x = \int \langle \rho E_x \rangle d^3 r + \frac{1}{c} \int \langle \mathbf{j} \times \mathbf{H} \rangle_x d^3 r = \langle \nabla_x (\mathbf{d} \cdot \mathbf{E} + \mathbf{m} \cdot \mathbf{H}) \rangle. \tag{5}$$

Up to this point formulas (1)-(5) coincide with those in [23], where the particle has no own rotation. However, all physical values depend implicitly on the angular velocity $\mathbf{\Omega}$, while the particle experiences the action of torque $\mathbf{M} = (M_x, M_y, M_z)$:

$$\mathbf{M} = \langle (\mathbf{d} \times \mathbf{E} + \mathbf{m} \times \mathbf{H}) \rangle \tag{6}$$

The work of the fluctuating field in $\Sigma'$ (Eq. (3)) is spent on stopping particle rotation and its heating,

$$\int_{\Omega'} \langle \mathbf{j}' \cdot \mathbf{E}' \rangle d^3 r' = \langle \dot{\mathbf{d}}' \mathbf{E}' + \dot{\mathbf{m}}' \mathbf{H}' \rangle = \dot{Q}'' + M'_x \Omega \cos\theta + M'_y \Omega \sin\theta, \tag{7}$$

where $\dot{Q}'' = dQ''/dt'' = d(C_0 T_1)/dt''$ is the particle heating rate in its rest frame $\Sigma''$, $C_0$ is the particle heat capacity. With allowance for the latter relation, (3), (7), relativistic transformations of torque $M'_x = \gamma M_x$, $M'_y = M_y$, $M'_z = M_z$ and time $dt'' = dt' = \gamma^{-1} dt$ ($\gamma = (1-\beta^2)^{-1/2}$ is the Lorentz-factor) one obtains

$$d(C_0 T_1)/dt = \gamma \, dQ/dt - \Omega \left( M_x \cos\theta + M_z \gamma^{-1} \sin\theta \right). \tag{8}$$

Eq. (8) allows one to trace the evolution of particle's proper temperature according to the time in $\Sigma$. It should be noted that the component $M_y$ of torque does not perform the work and may only cause the precession of particle.

Eqs. (4)–(8) are the basis for further analysis. The calculations in the right-hand sides of (4)–(6) are performed using our method [9], by representing the dipole moments and fields as the sums of spontaneous and induced components and calculating their pair correlation products with the use of fluctuation-dissipation relations. The required fluctuation-dissipation relations of dipole moments in $\Sigma''$ are modified due to particle rotation. The elements of general transformation matrix relating the components of vectors in $\Sigma'$ and $\Sigma''$ are given by [29]

$$A_{ik} = n_i n_k + (\delta_{ik} - n_i n_k)\cos\Omega t' - e_{ikl} n_l \sin\Omega t', \tag{9}$$

where $n_i = (\cos\theta, \sin\theta, 0)$ are the components of the unit vector **n** in the direction of angular velocity (Fig. 1). By setting the condition $R\Omega/c \ll 1$ one obtains $t' = t''$ in $\Sigma'$ and $\Sigma''$.

Using (9), the fluctuation-dissipation relations for spontaneous and induced dipole moments in co-moving frame $\Sigma'$ take the form

$$\left\langle d_x^{sp'}(\omega) d_x^{sp'}(\omega') \right\rangle = \frac{1}{2} 2\pi\hbar\delta(\omega+\omega') \cdot$$
$$\cdot \left\{ 2\cos^2\theta\, \alpha_e''(\omega)\coth\frac{\hbar\omega}{2k_B T_1} + \sin^2\theta \left[ \alpha_e''(\omega^+)\coth\frac{\hbar\omega^+}{2k_B T_1} + \alpha_e''(\omega^-)\coth\frac{\hbar\omega^-}{2k_B T_1} \right] \right\} \tag{10}$$

$$\left\langle d_z^{sp'}(\omega) d_z^{sp'}(\omega') \right\rangle = \frac{1}{2} 2\pi\hbar\delta(\omega+\omega') \cdot$$
$$\cdot \left\{ 2\sin^2\theta\, \alpha_e''(\omega)\coth\frac{\hbar\omega}{2k_B T_1} + \cos^2\theta \left[ \alpha_e''(\omega^+)\coth\frac{\hbar\omega^+}{2k_B T_1} + \alpha_e''(\omega^-)\coth\frac{\hbar\omega^-}{2k_B T_1} \right] \right\} \tag{11}$$

$$\left\langle d_y^{sp'}(\omega) d_y^{sp'}(\omega') \right\rangle = \frac{1}{2} 2\pi\hbar\delta(\omega+\omega') \cdot \left[ \alpha_e''(\omega^+)\coth\frac{\hbar\omega^+}{2k_B T_1} + \alpha_e''(\omega^-)\coth\frac{\hbar\omega^-}{2k_B T_1} \right] \tag{12}$$

$$\left\langle d_x^{sp'}(\omega) d_y^{sp'}(\omega') \right\rangle = -\left\langle d_y^{sp'}(\omega) d_x^{sp'}(\omega') \right\rangle = \frac{i}{2}\sin\theta \cdot 2\pi\hbar\delta(\omega+\omega') \cdot$$
$$\cdot \left[ \alpha_e''(\omega^+)\coth\frac{\hbar\omega^+}{2k_B T_1} - \alpha_e''(\omega^-)\coth\frac{\hbar\omega^-}{2k_B T_1} \right] \tag{13}$$

$$\left\langle d_y^{sp'}(\omega) d_z^{sp'}(\omega') \right\rangle = -\left\langle d_z^{sp'}(\omega) d_y^{sp'}(\omega') \right\rangle = \frac{i}{2}\cos\theta \cdot 2\pi\hbar\delta(\omega+\omega') \cdot$$
$$\cdot \left[ \alpha_e''(\omega^+)\coth\frac{\hbar\omega^+}{2k_B T_1} - \alpha_e''(\omega^-)\coth\frac{\hbar\omega^-}{2k_B T_1} \right] \tag{14}$$

$$\left\langle d_x^{sp'}(\omega) d_z^{sp'}(\omega') \right\rangle = \left\langle d_z^{sp'}(\omega) d_x^{sp'}(\omega') \right\rangle = \sin\theta\sin\theta \cdot 2\pi\hbar\delta(\omega+\omega')$$
$$\left[ \alpha_e''(\omega)\coth\frac{\hbar\omega}{2k_B T_1} - \frac{1}{2}\left( \alpha_e''(\omega^+)\coth\frac{\hbar\omega^+}{2k_B T_1} + \alpha_e''(\omega^-)\coth\frac{\hbar\omega^-}{2k_B T_1} \right) \right] \tag{15}$$

where $\omega^{\pm} = \omega \pm \Omega$, $\alpha''_e$ is the imaginary component of the electric polarizability. The same expressions are valid for the magnetic dipole moments with replacing $\alpha_e \to \alpha_m$. It is worth noting that fluctuation-dissipation relations (10)–(15) are more general than those used in [16, 24].

### 3. Frictional force, heating rate and components of torque

Using (4)–(6) and (10)–(15), the calculations are performed in line with [23, 24]. The resulting expressions for $F_x, dQ/dt, M_x, M_z$ have the form

$$F_x = -\frac{\hbar \gamma}{4\pi c^4} \int_{-\infty}^{+\infty} d\omega \omega^4 \int_{-1}^{1} dx\, x \cdot$$
$$\left[ \begin{array}{l} \alpha''(\omega_\beta) f_1(x,\beta,\theta) \left( \coth \frac{\hbar \omega}{2k_B T_2} - \coth \frac{\hbar \omega_\beta}{2k_B T_1} \right) + \\ + \alpha''(\omega_\beta^+) f_2(x,\beta,\theta) \left( \coth \frac{\hbar \omega}{2k_B T_2} - \coth \frac{\hbar \omega_\beta^+}{2k_B T_1} \right) \end{array} \right] \quad (16)$$

$$\dot{Q} = \frac{\hbar \gamma}{4\pi c^3} \int_{-\infty}^{+\infty} d\omega \omega^4 \int_{-1}^{1} dx (1+\beta x) \cdot$$
$$\cdot \left\{ \begin{array}{l} f_1(x,\beta,\theta) \cdot \alpha''(\omega_\beta) \left( \coth \frac{\hbar \omega}{2k_B T_2} - \coth \frac{\hbar \gamma \omega_\beta}{2k_B T_1} \right) + \\ f_2(x,\beta,\theta) \cdot \alpha''(\omega_\beta^+) \left( \coth \frac{\hbar \omega}{2k_B T_2} - \coth \frac{\hbar \gamma \omega_\beta^+}{2k_B T_1} \right) \end{array} \right\} \quad (17)$$

$$M_x = -\frac{\hbar \gamma \cos\theta}{4\pi c^3} \int_{-\infty}^{+\infty} d\omega \omega^3 \int_{-1}^{1} dx\, \alpha''(\omega_\beta^+) \left[(1+x^2)(1+\beta^2) + 4\beta x\right] \left( \coth \frac{\hbar \omega}{2k_B T_2} - \coth \frac{\hbar \omega_\beta^+}{2k_B T_1} \right) \quad (18)$$

$$M_z = -\frac{\hbar \sin\theta}{8\pi c^3} \int_{-\infty}^{+\infty} d\omega \omega^3 \int_{-1}^{1} dx\, \alpha''(\omega_\beta^+)(3 - x^2 + 2\beta x) \left( \coth \frac{\hbar \omega}{2k_B T_2} - \coth \frac{\hbar \omega_\beta^+}{2k_B T_1} \right) \quad (19)$$

where $\alpha'' = \alpha''_e + \alpha''_m$, $\omega_\beta = \gamma \omega (1+\beta x), \omega_\beta^+ = \gamma \omega (1+\beta x) + \Omega$, and the auxiliary functions $f_i(x,\beta,\theta)$ ($i=1,2$) are given by

$$f_1(x,\beta,\theta) = (1-\beta^2)(1-x^2)\cos^2\theta + \left((1+\beta^2)(1+x^2) + 4\beta x\right) \frac{\sin^2\theta}{2} \quad (20)$$

$$f_2(x,\beta,\theta) = (1-\beta^2)(1-x^2)\sin^2\theta + \left((1+\beta^2)(1+x^2) + 4\beta x\right) \frac{1+\cos^2\theta}{2} \quad (21)$$

Eqs. (16)–(19) agree with the results [24] obtained at $\theta = 0, \pi/2$.

### 4. Intensity of radiation

Using (2) and (3), the power of radiation is written in the form

$$I = I_1 - I_2 = -\left(\frac{dQ}{dt} + F_x V\right). \tag{22}$$

Substituting (15) and (16) into (18) yields

$$I = I_1(T_1) - I_2(T_2) = \frac{\hbar \gamma}{4\pi c^3} \int_{-\infty}^{+\infty} d\omega \omega^4 \int_{-1}^{1} dx \cdot \begin{bmatrix} \alpha''(\omega_\beta) f_1(x,\beta,\theta)\left(\coth\frac{\hbar\omega_\beta}{2k_B T_1} - \coth\frac{\hbar\omega}{2k_B T_2}\right) + \\ + \alpha''(\omega_\beta^+) f_2(x,\beta,\theta)\left(\coth\frac{\hbar\omega_\beta^+}{2k_B T_1} - \coth\frac{\hbar\omega}{2k_B T_2}\right) \end{bmatrix} \tag{23}$$

A very important consequence of (23) is the existence of nonthermal radiation in the system "particle-vacuum" at $\Omega \neq 0$. By performing the limiting transition $T_1 \to 0, T_2 \to 0$ in (23) and (15) one obtains

$$I^{(0)} \equiv I_1(0) = \frac{\hbar\gamma}{2\pi c^3}\int_{-1}^{1} dx \cdot f(x,\theta) \int_{0}^{\Omega\gamma^{-1}(1+\beta x)^{-1}} d\omega \omega^4 \alpha''(\Omega - \gamma\omega(1+\beta x)) =$$

$$= \frac{4\hbar}{3\pi c^3}\int_{0}^{\Omega} d\xi \xi^4 \alpha''(\Omega - \xi) \tag{24}$$

The last line of Eq. (20) agrees with [20] in the case of rotating body without translational motion. However, as it stems from (23) and (24), though the integral radiation power is $\gamma$ – independent, the spectral-angular distribution significantly depends on relativistic factor and mutual angular orientation of the linear and angular velocity vectors.

By performing the replacement $\theta \to \theta_0$ in (24) (not to be confused with the photon angle $\theta$ with respect to the direction of **V**), and taking into account that $x \equiv -\cos\theta$, the spectral-angular intensity of radiation per unit solid angle $d\tilde{\Omega} = 2\pi\sin\theta\,d\theta$ takes the form

$$\frac{d^2 I}{d\omega d\tilde{\Omega}} = \frac{\gamma\hbar\omega^4}{4\pi^2 c^3}\Theta(\Omega - \gamma\omega(1-\beta\cos\theta))\alpha''((\Omega - \gamma\omega(1-\beta\cos\theta))\cdot$$

$$\cdot\left[(1-\beta^2)(1-\cos^2\theta)\sin^2\theta_0 + \left((1+\beta^2)(1+\cos^2\theta) - 4\beta\cos\theta\right)\frac{1+\cos^2\theta_0}{2}\right] \tag{25}$$

where $\Theta(x)$ – the Heaviside step-function.

From (25) it follows that nonthermal radiation is generated in the frequency range $0 < \omega < \dfrac{\Omega\sqrt{1-\beta^2}}{(1-\beta\cos\theta)}$, and the maximum frequency $\omega_{max} = \Omega\sqrt{\dfrac{1+\beta}{1-\beta}}$ is emitted in the direction of particle motion, $\theta = 0$. In the opposite direction, the frequency of radiation is $\omega_{min} = \Omega\sqrt{\dfrac{1-\beta}{1+\beta}}$. At $\beta \ll 1$. The spectral-angular intensity does not depend on the linear velocity and has the simplest form

$$\frac{d^2 I}{d\omega d\tilde{\Omega}} = \frac{\hbar\omega^4}{4\pi^2 c^3}\Theta(\Omega-\omega)\alpha''(\Omega-\omega)\left(\sin^2\theta\sin^2\theta_0 + (1+\cos^2\theta)\frac{1+\cos^2\theta_0}{2}\right). \quad (26)$$

On the whole, as we can see from (25) and (26), the shape of the spectrum is determined by the dielectric properties of particle.

The phenomenon of nonthermal radiation from rotating particle is intimately related to *superradiance* first discovered by Zel'dovich [3] and, according to the interpretation given in [20–22, 26], one may speak about quantum instability of vacuum in the considered system that is analogous to the process of electron-positron birth in strong electric fields and Hawking's radiation in strong gravitation field.

### 5. Particle dynamics and heating rate at zero temperature conditions

The dynamics of translational and rotational motion in the case $T_1 \to 0, T_2 \to 0$ can be analyzed using Eqs. (16)–(19). Having performed the corresponding limiting transitions and integrations over variable $x$ one obtains

$$F_x^{(0)} = -\frac{4\hbar V}{3\pi c^5}\int_0^\Omega d\xi \xi^4 \alpha''(\Omega-\xi), \quad (27)$$

$$\frac{dQ}{dt} = -\frac{4\hbar}{3\pi c^3 \gamma^2}\int_0^\Omega d\xi \xi^4 \alpha''(\Omega-\xi), \quad (28)$$

$$M_x = -\frac{4\hbar\cos\theta}{3\pi c^3 \gamma}\int_0^\Omega d\xi \xi^3 \alpha''(\Omega-\xi), \quad (29)$$

$$M_z = -\frac{4\hbar\sin\theta}{3\pi c^3}\int_0^\Omega d\xi \xi^3 \alpha''(\Omega-\xi). \quad (30)$$

Using (22), (24), (27) and (28) yields

$$F_x^{(0)} = -\frac{\beta}{c} \cdot I^{(0)}. \quad (31)$$

Therefore, it is the force $F_x^{(0)}$ acting on the particle in the frame of reference of vacuum background determines its nonthermal radiation.

The particle acceleration is determined from the dynamic equation

$$\frac{d}{dt}\left(\frac{mV}{\sqrt{1-V^2/c^2}}\right) = F_x. \qquad (32)$$

Following [27], Eq. (32) can be written in the form

$$\gamma^3 m \frac{dV}{dt} = F_x - \frac{\beta}{c}\gamma^2 \frac{dQ}{dt}. \qquad (33)$$

Equation (33) takes into account the required derivative $dm/dt$ expressed via $dQ/dt$. Substituting (27) and (28) into (33) yields

$$\frac{d\beta}{dt} = 0. \qquad (34)$$

Therefore $\beta = const$ in this case.

To examine rotational dynamics of particle, we write the corresponding dynamic equations in co-moving coordinate system $\Sigma'$

$$I_{ik}\frac{d\Omega_k}{dt'} = M'_k. \qquad (35)$$

where $I_{ik}$ – the components of inertia moment tensor in $\Sigma'$ ($I_{ik} = I_0 \delta_{ik}$ for spherical particle in $\Sigma'$), and the projections $M'_k$ of torque should be expressed via the projections of $M_k$ in $\Sigma$. Moreover, we introduce the projections $M'_n$ and $M'_\perp$ of torque onto the direction of particle rotation axis and the perpendicular direction in the plane ($x', z'$), namely

$$M'_n = M'_x \cos\theta + M'_z \sin\theta, \quad M'_\perp = -M'_x \sin\theta + M'_z \cos\theta \qquad (36)$$

Using these definitions of $M'_n$ and $M'_\perp$ Eqs. (35) take the form

$$I_0 \frac{d\Omega}{dt'} = M'_n, \qquad (37)$$

$$I_0 \Omega \frac{d\theta}{dt'} = M'_\perp. \qquad (38)$$

Furthermore, with allowance for (36) and relativistic transformations $M'_x = \gamma M_x$, $M'_z = M_z$, $dt' = dt/\gamma$, Eqs. (37) and (38) are rewritten in terms of $M_x$ and $M_z$

$$I_0 \frac{d\Omega}{dt} = M_x \cos\vartheta + \gamma^{-1} M_z \sin\theta ,\qquad(37a)$$

$$I_0 \Omega \frac{d\theta}{dt} = -M_x \sin\theta + \gamma^{-1} M_z \cos\theta .\qquad(38a)$$

Substituting (29), (30) into (37a), (38a) yields

$$I_0 \frac{d\Omega}{dt} = -\frac{4\hbar}{3\pi c^3} \int_0^\Omega d\xi \xi^3 \alpha''(\Omega - \xi) ,\qquad(39)$$

$$\frac{d\theta}{dt} = 0 .\qquad(40)$$

From (39) and (40) we see that the particle monotonously slows down at $\theta = const$, i. e. the mutual orientation of angular and linear velocity vectors does not change with time. A possibility of particle precession around the direction of **V** requires further investigation, but it does not affect the intensity of radiation and other quantities calculated here and in what follows.

To analyze the kinetics of particle heating, we substitute Eqs. (29), (30) into (8) and obtain

$$\frac{d(C_0 T_1)}{dt} = \frac{4\hbar}{3\pi c^3 \gamma} \int_0^\Omega d\xi \xi^3 (\Omega - \xi) \alpha''(\Omega - \xi) .\qquad(41)$$

Eq. (41) shows that the particle temperature increases with time, depending on the behavior of $C_0$ near the state $T_1 = 0$. Therefore, the state $T_1 = 0$ is unstable, and the particle is heated due to the dissipation of the kinetic energy of rotation. Another part of this kinetic energy is radiated. At finite temperature of particle and (or) background, the radiation intensity contains both nonthermal and thermal contributions (the latter being prevailing).

### 6. Case study: conducting particle

*6.1 General equations*

In order to simplify the analysis, we adopt a simple form of the dielectric polarizability of spherical nonmagnetic particle: $\alpha''(\omega) = 3R^3 \omega / 4\pi \sigma_0$, where $\sigma_0$ is the static conductivity. This corresponds to the law-frequency limit of the Drude dielectric permittivity $\varepsilon(\omega) = i \cdot 4\pi \sigma_0 / \omega$. The effects of magnetic polarizability can be considered numerically (see, [24], for example). Using (15)–(19), (23) with this simple form of polarizability, the integrals are calculated explicitly (see Appendix A) and one obtains

$$F_x = -\frac{a\beta}{c}\left[\gamma^2(1+\beta^2/5)\vartheta_2^6 + \psi_1(\Omega/\vartheta_1)\vartheta_1^6\right], \quad (42)$$

$$dQ/dt = a\left[\gamma^2(1+2\beta^2+\beta^4/5)\vartheta_2^6 - \gamma^{-2}\psi_1(\Omega/\vartheta_1)\vartheta_1^6\right], \quad (43)$$

$$I \equiv I_1 - I_2 = a\left[\vartheta_1^6 \psi_1(\Omega/\vartheta_1) - \gamma^2(1+\beta^2)\vartheta_2^6\right], \quad (44)$$

$$M_x = -\frac{7}{20\pi^2}a\Omega\cos\theta\left[\gamma(1+\beta^2)\vartheta_2^4 + \gamma^{-1}\psi_2(\Omega/\vartheta_1)\vartheta_1^4\right] \quad (45)$$

$$M_z = -\frac{7}{20\pi^2}a\Omega\sin\theta\left[\vartheta_2^4 + \psi_2(\Omega/\vartheta_1)\vartheta_1^4\right] \quad (46)$$

$$\psi_1(x) = 1 + \frac{21}{10\pi^2}x^2 + \frac{7}{8\pi^4}x^4 + \frac{7}{80\pi^6}x^6, \quad (47)$$

$$\psi_2(x) = 3 + \frac{5}{2\pi^2}x^2 + \frac{3}{8\pi^4}x^4, \quad (48)$$

where $a = \dfrac{8\pi^4}{21}\dfrac{\hbar R^3}{c^3 \sigma_0}$, $\vartheta_{1,2} = k_B T_{1,2}/\hbar$.

Substituting (43), (45) and (46) into (8) yields

$$\frac{d(C_0 T_1)}{dt} = a\left[\gamma^3(1+2\beta^2+\beta^4/5)\vartheta_2^6 - \gamma^{-1}\psi_1(\Omega/\vartheta_1)\vartheta_1^6\right] +$$
$$+ \frac{7}{20\pi^2}a\Omega^2\left[\left(\gamma(1+\beta^2)\cos^2\theta + \gamma^{-1}\sin^2\theta\right)\vartheta_2^4 + \gamma^{-1}\psi_2(\Omega/\vartheta_1)\vartheta_1^4\right] \quad (49)$$

Further on, taking into account (42) and (43), dynamic equation (33) takes the form

$$\frac{d\beta}{dt} = -\frac{a\vartheta_2^6}{mc^2}\beta\left[\frac{1+2\beta^2+\beta^4/5}{\sqrt{1-\beta^2}} + \sqrt{1-\beta^2}(1+\beta^2/5)\right]. \quad (50)$$

As follows from (50), the dynamics of translational motion is independent of $\Omega$. Finally, substituting (46) and (47) into (37a) and (38a) yields

$$I_0 \frac{d\Omega}{dt} = -\frac{7}{20\pi^2}a\Omega\left[\left(\gamma(1+\beta^2)\cos^2\theta + \gamma^{-1}\sin^2\theta\right)\vartheta_2^4 + \gamma^{-1}\psi_2(\Omega/\vartheta_1)\vartheta_1^4\right], \quad (51)$$

$$I_0 \frac{d\theta}{dt} = \frac{7}{10\pi^2}a\gamma\beta^2\vartheta_2^4\sin\theta\cos\theta. \quad (52)$$

A simple relation between $\Omega$ and $\theta$ can be obtained in the most realistic case $\Omega/\vartheta_1 < 1$. Then $\psi_2(\Omega/\vartheta_1) \approx 3$ and from (51), (52) one obtains

$$\frac{d\Omega}{\Omega} = -\frac{\left[3(T_1/T_2)^4 \gamma^{-2} + (1+\beta^2)\cos^2\theta + \gamma^{-2}\sin^2\theta\right]}{2\beta^2 \sin\theta\cos\theta} d\theta. \qquad (53)$$

Integrating Eq. (53) yields

$$\frac{\Omega}{\Omega_0} = \left(\frac{\sin\theta_0}{\sin\theta}\right)^p \left(\frac{\cos\theta}{\cos\theta_0}\right)^q, \qquad (54)$$

$$p = \frac{1+\beta^2}{2\beta^2} + \frac{3}{2\beta^2\gamma^2}\left(\frac{T_1}{T_2}\right)^4, \qquad (55)$$

$$q = \frac{1}{2\beta^2\gamma^2} + \frac{3}{2\beta^2\gamma^2}\left(\frac{T_1}{T_2}\right)^4. \qquad (56)$$

where $\Omega_0, \theta_0$ are the initial values of $\Omega$ and $\theta$ at $t=0$, $p=(1+\beta^2)/2\beta^2$, $q=3(T_1/T_2)^4/(2\beta^2\gamma^2)$, and $r=1/(2\beta^2\gamma^2)$. Since from (51) it follows that $\Omega$ monotonously goes to zero, the right-hand side of (54) also goes to zero. This means that $\theta \to \pi/2$, i. e the angular velocity vector turns perpendicularly to the vector of linear velocity.

### 6.2 Time scales of heating and deceleration

The characteristic time-scales of heating and deceleration can be obtained when writing Eqs. (49), (50) and (51) in dimensionless form. We make the definitions $C_0 = 4\pi R^3 \rho C_s$, $m = 4\pi R^3 \rho/3$, and $I_0 = 2mR^2/5 = 8\pi\rho R^5/15$ where $\rho$ and $C_s$ are the density of mass and the specific heat capacity (assuming that $C_s = const$). Then the corresponding time scales of thermal relaxation $\tau_Q$, linear $\tau_\beta$ and angular $\tau_\Omega$ deceleration are (see Appendix B)

$$\tau_Q = \frac{7}{2\pi^3} \frac{c^3 \sigma_0 C_s \rho}{k_B \vartheta_2^5}, \qquad (57)$$

$$\tau_\beta = \frac{7}{2\pi^3} \frac{c^5 \sigma_0 \rho}{\hbar \vartheta_2^6}, \qquad (58)$$

$$\tau_\Omega = \frac{4}{\pi} \frac{c^3 \sigma_0 \rho R^2}{\hbar \vartheta_2^4}. \qquad (59)$$

Simple analysis of (57)–(59) shows that $\tau_Q \ll \tau_\Omega \ll \tau_\beta$ for conducting nanoparticles in a wide range of temperatures. For example, for graphite particles at $T_2 = 300\,K$, $\sigma_0 = 2\cdot 10^{14}\,s^{-1}$ $C_s = 7\cdot 10^6\,erg/g\cdot K$, $\rho = 2.1\,g/cm^3$, $R = 10\,nm$ we obtain $\tau_Q = 0.7\,s$, $\tau_\Omega = 5.7\cdot 10^6\,s$, and $\tau_\beta = 2.9 \cdot 10^{11} s$. Therefore, when calculating temporal dependence of $T_1/T_2$ from (49), we may fix the parameters $\Omega, \beta$. On the other hand, when calculating temporal dependences of $\Omega$ and $\theta$ we may fix the ratio $T_1/T_2$ and $\beta$.

## 6.3 Numerical results

As follows from (50), the dynamics of linear motion is independent of particle temperature and the angular velocity. Eq. (50) is integrated with the result

$$f(\beta_0) - f(\beta) = (t - t_0)/\tau_\beta \tag{58}$$

$$f(\beta) = -\frac{1}{2} arth\left(\frac{1}{\sqrt{1-\beta^2}}\right) + \sqrt{\frac{2}{3}} arth\left(\sqrt{\frac{3(1-\beta^2)}{8}}\right), \tag{59}$$

where $\beta_0 = \beta(t_0)$. The linear dynamics can be analyzed at fixed values of $T_2$. According to (59), parameter $\beta$ asymptotically changes from $\beta_0$ to 0, and one can find only the time $\tau_s$ required for deceleration from $\beta_0$ to a certain value $\beta < \beta_0$. Fig. 2 shows the ratio $\tau_s/\tau_\beta$ depending on $\beta_0$ and assuming that $\tau_s$ is the time of deceleration from $\beta_0$ to $10^{-4}\beta_0$.

At fixed $T_2$, the intensity of radiation is determined by $T_1$, while the dependences on the linear and angular velocities influence in an indirect way through $T_1$. Irrespectively of the initial ratio between $T_1$ and $T_2$, the temperature $T_1$ quickly tends to quasistationary value, so does the intensity of radiation.

It is worth noting two important cases are when $T_1 \neq T_2$ (with $T_1$ being the initial particle temperature and $T_2 = const$): i) $t < \tau_s$; ii) $t > \tau_s$.

At $t < \tau_s$ and $T_1 < T_2$ (even at $T_2 = 1K$, except the case $\Omega > \vartheta_2$, where $\Omega > \theta_2 = 10^{11} s^{-1}$), the absorbed power $I_2$ is much greater than the emitted power $I_1$ (compare the second and first terms in (44)). To an outside observer, for example, a cloud of cold dust particles being injected into the area with a higher temperature of background radiation will effectively absorb thermal photons, and the absorbed power will scale as $\gamma^2$. The radiated power (the second term in (44)) in this case depends on the ratio between $\Omega$ and $\vartheta_1$. The function $\vartheta_1^6 \psi_1(\Omega/\vartheta_1)$ reaches a maximum value at $\Omega/\vartheta_1 = 4.78$, and the maximum of $I_1$ is an order of magnitude higher than the intensity $I^{(0)}$ of nonthermal radiation (Eq. (24)). At $t < \tau_s$ and $T_1 > T_2$ the initial ratio $I_1/I_2$ strongly depends on $\beta$ and can be both $<1$ (at $\gamma \gg 1$) and $>1$ (at $\gamma \to 1$).

At $t > \tau_s$, the temperature reaches a quasistationary value which is determined by equating the right hand side of (49) to zero. In this case the ratio $I_1/I_2$ between the emitted and absorbed power exceeds the unit. It scales as $\gamma^2$ and increases with $\Omega/\vartheta_2$. In the case when the temperature of particle reaches the melting point, a special analysis is needed.

The temporal dependences of $T_1/T_2$ and $I_1/I_2$ for graphite particles are shown in Figs. 3, 4 assuming that $T_2 = const$. Solid and dashed lines correspond to $\beta = 0.1$ and $\beta = 0.999$. Thin lines (solid and dashed) correspond to the initial ratio $T_1/T_2 = 3$, while thick lines (solid and dashed) correspond to the initial ratio $T_1/T_2 = 0.01$. It should be noted that the time scale in Fig. 3 is given in units of $\tau_Q$, while in Fig. 4 – in units of $\tau_{ST}$ (individual for each curve), where $\tau_{ST}$ corresponds to the onset of saturation of curves shown in Fig. 3. In all these cases we assumed that $\Omega/\theta_1 = 1$ and $\theta = \pi/2$. One can see that $T_1$ and $I_1$ quickly go to quasistationary values. At $t \gg \tau_s$, parameters $\beta$ and $\Omega$ monotonously decrease with time and $I_1 \to I_2$ until the particle stopping.

## 7. Conclusions

We have obtained a full set of equations describing the fluctuation-electromagnetic interaction, dynamics and kinetics of heating of a small polarizable particle with arbitrary direction of spin moving in a vacuum background at arbitrary temperatures of particle and background. The obtained equations allow one to calculate the temporal evolution of thermal and dynamic state of particle.

At zero temperature of particle and radiation background, the particle emits long-wavelength photons. The orientation of the angular velocity vector with respect to the linear velocity vector is independent of time. The corresponding intensity of radiation depends on the angular velocity and does not depend on the linear velocity and spin direction. The linear velocity is constant, but since the particle is heated, the state $T_1 = 0$ is unstable.

We have also analyzed the case of finite temperatures of particle and background assuming that the particle polarizability is proportional to frequency. As in the "cold" case, the linear acceleration of the particle does not depend on its temperature and angular velocity. The power of radiation depends on the angular and linear velocities. The time dependence of the complete set of dynamic parameters is determined by the local particle temperature (in its rest frame), since the time of thermal relaxation is much less than the times of stopping (for both rotational and translational motion). The particle temperature relaxes to a quasistationary value depending on the background radiation temperature, the linear and angular velocities. The spin of particle tends to be oriented perpendicularly to the linear velocity. The maximum radiated power in a quasistationary state is proportional to $\gamma^2$. Further on, the difference between the emitted and absorbed intensities of radiation decreases with time until the particle stops. The initially cold particle predominantly absorbs background radiation in the phase of heating and the rate of absorption is also proportional to $\gamma^2$. We believe that along with their fundamental importance, our results will be important to study and control nanoparticles trapped in cavities and in astrophysics of cosmic dust matter.

# References

[1] H.B.G. Casimir, Proc. K. Ned. Akad. Wet. 51 (1948) 793.

[2] D.A.R. Dalvit, P.A. Maia Neto, and F. D. Mazzitelli, in: Lecture Notes on "Casimir Physics" (ed. by D. A. R. Dalvit, P. Milonni, D. Roberts, and F da Rosa, Springer-Verlag, 2011). Vol. 834 (2011) 419; arXiv: 1101.2238.

[3] Ya. B. Zel'dovich, JETP. Lett. 14 (1971) 180.

[4] A. Einstein, Physik Zeitschr. 18 (1917) 121.

[5] G.V. Dedkov G V and A.A. Kyasov, Phys. Solid State 44/10(2002) 1809.

[6] V.A. Mkrtchian, V.A. Parsegian, R. Podgornik and W.M. Saslow, Phys. Rev. Lett. 91 (2003) 220801.

[7] J.R. Zurita-Sanchez, J.J. Greffet, and L. Novotny, Phys. Rev. A69 (2004) 022902.

[8] F. Intravaia, C. Henkel, and M. Antezza, in: Lecture Notes on "Casimir Physics" (ed. by D. A. R. Dalvit, P. Milonni, D. Roberts, and F da Rosa, Springer-Verlag, 2011). Vol. 834 (2011) 345.

[9] G.V. Dedkov and A.A. Kyasov, J. Phys.: Condens. Matter 20 (2008) 354006.

[10] S.Y. Buchmann and D. Welsch, Phys. Rev. A77 (2008) 012110.

[11] G. Barton, New J. Phys. 12(2010) 113045.

[12] K.A. Milton K A, J.S. Høye, and I. Brevik I Symmetry 8 (2016) 29.

[13] J.S. Høye, I. Brevik, K.A. Milton, J. *Phys. A: Math. Theor.* 48 (2015) 365004.

[14] J.S. Høye and I. Brevik, J. Phys.: Condens. Matter 27 (2015) 214008.

[15] G.V. Dedkov, A.A. Kyasov, Eur. Phys. Lett. 99 (2012) 64002

[16] R. Zhao, A. Manjavacas, F.G. Garcia de Abajo, and J.B. Pendry, Phys. Rev. Lett. 109 (2012) 2123604.

[17] Xiang Chen, Int. J. Mod. Phys. 27(14) (2013) 1350066; Int. J. Mod. Phys. 28(15) (2014)

1492002.

[18] V. Ameri, M.S. Aporvani, and F. Kheirandish, Phys. Rev. A92 (2015) 022110.

[19] A. Manjavacas and F. J. Garcia de Abajo, Phys. Rev. A82 (2010) 063827; Phys. Rev. Lett. 105 (2010) 113601.

[20] M.F. Maghrebi, R.L. Jaffe, and M. Kardar, Phys. Rev. Lett. 108 (2012) 230403 .

[21] M.F. Maghrebi, R. Golestanian, M. Kardar, Phys. Rev. A88 (2013) 230403.

[22] M.F. Maghrebi, R.L. Jaffe, and M. Kardar, Phys. Rev. A90 (2014) 012515.

[23] G.V. Dedkov, AA. Kyasov, Phys. Scripta, 89 (2014) 105501.

[24] A.A. Kyasov, G.V. Dedkov, Arm. J. Phys. 7/ 4 (2014)177.

[25] M. G. Silveirinha, New J. Phys. 16 (2014) 063011.

[26] G. Pieplow, C. Henkel, J. Phys.: Condens. Matter 27 (2015) 214001

[27] G.V. Dedkov, A.A. Kyasov, Int. J. Mod. Phys. 32 (2015) 1550237.

[28] A.I. Volokitin, B.N.J. Persson, Phys. Rev. B93 (2016) 035407.

[29] G.V. Dedkov, A.A. Kyasov, Tech. Phys. Lett. 42(1) (2016) 8.

[30] A.A. Kyasov, G.V. Dedkov, Phys. J. 2(3) (2016) 176.

**Appendix A**

All double integrals in (16)–(19) and (23) have the same structure and can be written in the similar form when using $\alpha''(\omega) = 3R^3\omega/(4\pi\sigma_0)$, denoting $\theta_{1,2} = k_B T_{1,2}/\hbar$ and $z = \gamma(1+\beta x)$. For example, Eq. (19) takes the form

$$M_x = -\frac{63a}{54\pi^2}\gamma\cos\theta\left[2\Omega I^{(1)} - \gamma^{-4}I^{(2)}\right] \tag{A1}$$

$$I^{(1)} = \int_0^\infty d\omega \frac{\omega^3}{\exp(\omega/\theta_2)-1}\int_{-1}^1 [(1+x^2)(1+\beta^2)+4\beta x]dx \tag{A2}$$

$$I^{(2)} = -\int_0^\infty dz\, z^3\left[\frac{z+\Omega}{\exp((z+\Omega)/\theta_1)-1} - \frac{z-\Omega}{\exp((z-\Omega)/\theta_1)-1}\right]\int_{-1}^1 dx\frac{(1+x^2)(1+\beta^2)+4\beta x}{(1+\beta x)^4} \tag{A3}$$

Integral (A2) is calculated with the result

$$I^{(1)} = \theta_2^4\frac{8}{3}(1+\beta^2)\int_0^\infty dx\frac{x^3}{(\exp(x)-1)} = \frac{8\pi^4}{45}\theta_2^4(1+\beta^2) \tag{A4}$$

The first integral in (A3) is

$$\int_0^\infty dz\, z^3\left[\frac{z+\Omega}{\exp((z+\Omega)/\theta_1)-1} - \frac{z-\Omega}{\exp((z-\Omega)/\theta_1)-1}\right] =$$

$$= -\theta_1^5\left[\int_0^\infty dx\frac{6x^2 b + 2xb^3}{\exp(x)-1} - \int_0^b dx\, x(x-b)^3\right] = -\theta_1^5\left(6b\frac{\pi^4}{15} + 2b^3\frac{\pi^2}{6} + \frac{b^5}{20}\right) = \tag{A5}$$

$$= -\frac{2\pi^4}{15}\Omega\theta_1^4\psi_2(b)$$

The second integral in (A3) is

$$\int_{-1}^1 dx\frac{(1+x^2)(1+\beta^2)+4\beta x}{(1+\beta x)^4} = \frac{8}{3}\gamma^2 \tag{A5}$$

where $b = \Omega/\theta_1$ and $\psi_2(x)$ is given by (48). Inserting (A2)–(A6) into (A1) yields (46). The needed intermediate integrals arising in calculating (42)–(43) and (45) are given by

$$\int_{-1}^1 dx\, x\frac{(1+x^2)(1+\beta^2)+4\beta x}{(1+\beta x)^5} = -\frac{8}{3}\beta\gamma^4 \tag{A7}$$

$$\int_{-1}^1 dx\, x\frac{(1-\beta^2)(1-x^2)}{(1+\beta x)^5} = -\frac{4}{3}\beta\gamma^4 \tag{A8}$$

$$\int_{-1}^1 dx\frac{(1-\beta^2)(1-x^2)}{(1+\beta x)^4} = \frac{4}{3}\gamma^2 \tag{A9}$$

$$\int_{-1}^1 dx\, x\frac{3-x^2+2\beta x}{(1+\beta x)^4} = \frac{16}{3}\gamma^4 \tag{A10}$$

$$\int_0^\infty dx \frac{x^5}{\exp(x)-1} = \frac{8\pi^6}{63} \tag{A11}$$

$$\int_0^\infty dz\, z^4 \left[ \frac{z+\Omega}{\exp((z+\Omega)/\theta_1)-1} + \frac{z-\Omega}{\exp((z-\Omega)/\theta_1)-1} \right] = \frac{16\pi^6}{63} \theta_1^6 \psi_1(b) \tag{A12}$$

where $\psi_1(x)$ is given by (47). Eq.(44) stems from (22), (42) and (43).

**Appendix B**

Eq. (49) reduces to dimensionless form by introducing the dimensionless variables of temperature $y = T_1/T_2$, angular velocity $x = \Omega/\theta_2$, and time $\tau = t/\tau_Q$. Taking into account that $C_0 T_1 = \frac{4\pi}{3} \rho R^3 C_s T_2 y$ we introduce $\tau_Q = \frac{7}{2\pi^3} \frac{\rho \sigma_0 c^3}{k_B \theta_2^5}$ and

$$dy/d\tau = \left[ \gamma^3(1+2\beta^2+\beta^4/5) - \frac{1}{\gamma}\psi_1\left(\frac{x}{y}\right) y^6 \right] +$$
$$+ \frac{7}{20\pi^2} x^2 \left[ \left( \gamma(1+\beta^2)\cos^2\theta + \frac{1}{\gamma}\sin^2\theta \right) + \frac{y^4}{\gamma}\psi_2\left(\frac{x}{y}\right) \right] \tag{B1}$$

The stationary temperature of particle is obtained from (A1) by equating the right hand side to zero. The expression for $\tau_\beta$ is obvious: from (50) it follows $\tau_\beta = \frac{mc^2}{a\theta_2^6} = \frac{7}{2\pi^3} \frac{\rho \sigma_0 c^5}{\hbar \theta_2^6}$.

Similarly to that, from (51) it follows $\tau_\Omega = \frac{20\pi^2}{7 I_0 a \theta_2^4} = \frac{4}{\pi} \frac{\rho \sigma_0 R^2}{\hbar \theta_2^4}$.

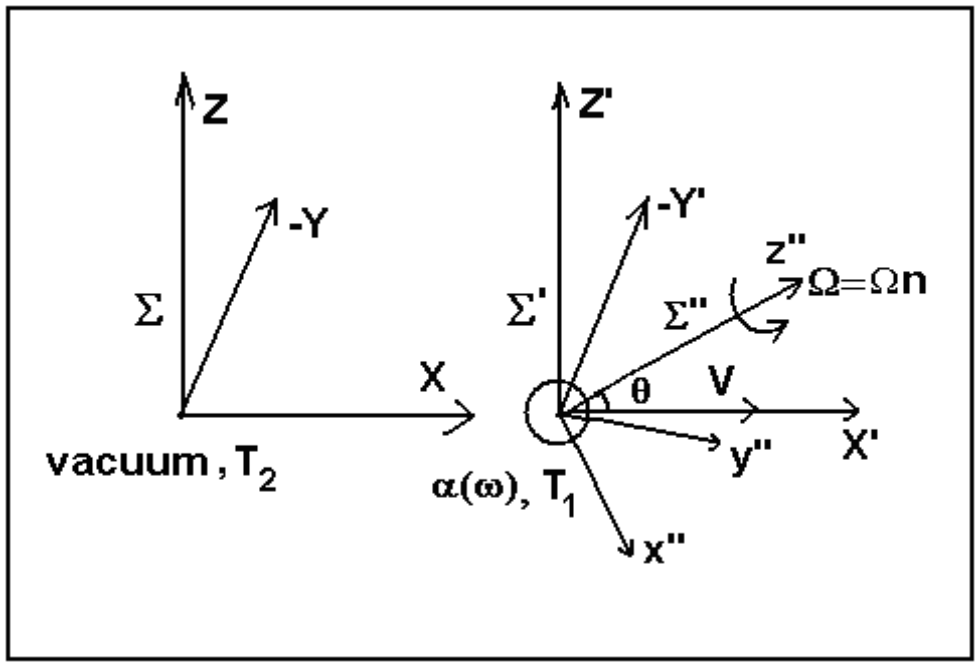

Fig. 1

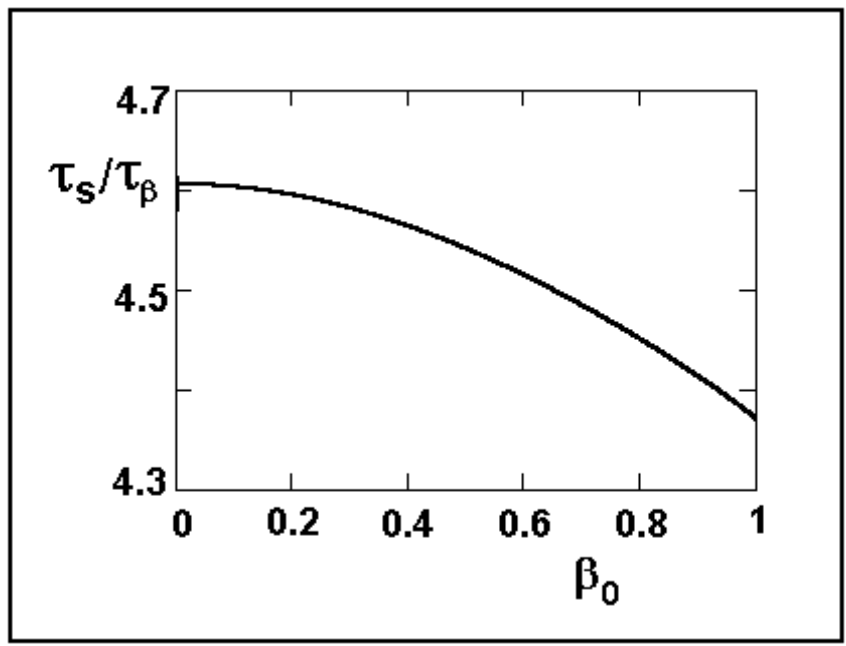

Fig. 2

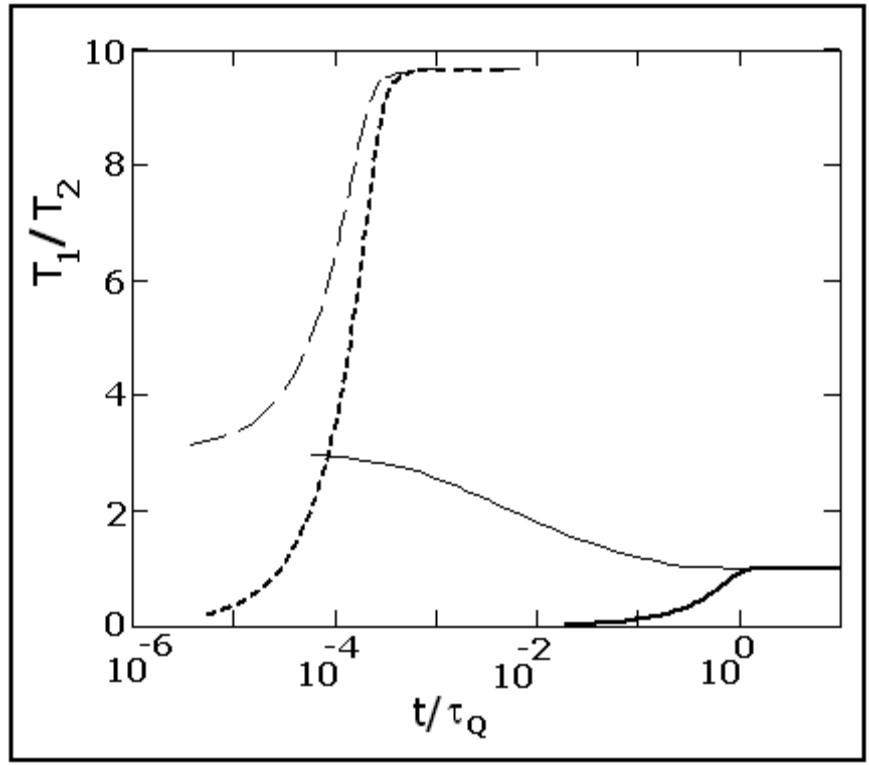

Fig. 3

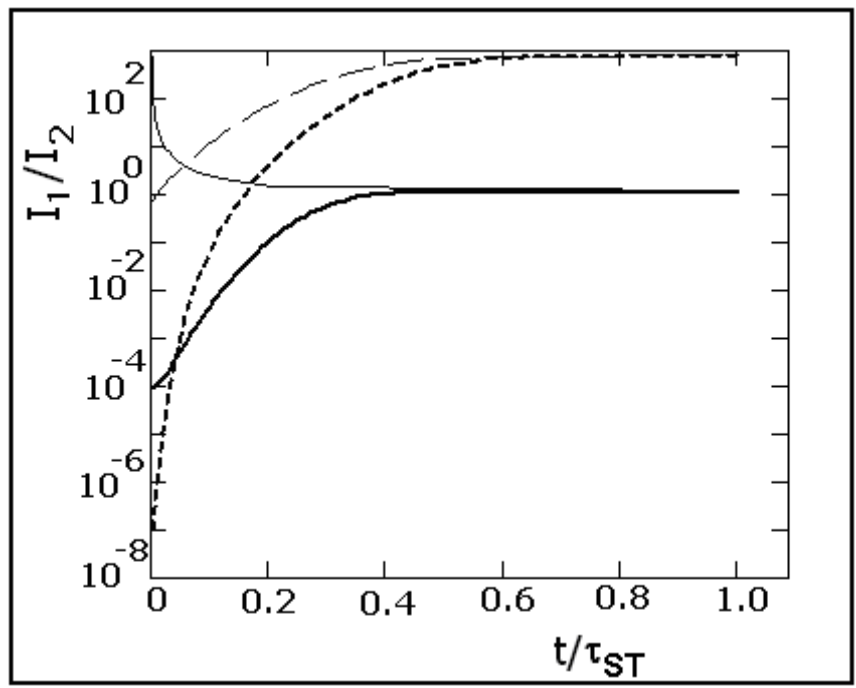

Fig. 4

# FIGURE CAPTIONS

Fig. 1

Scheme of the particle motion and coordinate systems used: $\Sigma$ is the reference frame of background radiation, $\Sigma'$ – reference frame co-moving with the particle (with velocity $V$), $\Sigma''$ – reference frame of particle rotating with angular velocity $\Omega$ with respect to $\Sigma'$. Vector **n** of the angular velocity direction lies in the plane $(x', z')$.

Fig. 2

The ratio $\tau_s / \tau_\beta$ depending on $\beta_0 = \beta(0)$. The stopping time $\tau_s$ corresponds to the particle deceleration from $\beta_0$ to $10^{-4} \beta_0$.

Fig. 3

Temporal dependence of $T_1/T_2$ according to Eq. (B1). In all the cases $T_2 = const$, $\Omega/\vartheta_2 = 1, \theta = \pi/2$, $\tau_Q$ is determined by (57). Thin solid line: $\beta = 0.1, T_1(0)/T_2 = 3$; thin dashed line: $\beta = 0.999, T_1(0)/T_2 = 3$; thick solid line: $\beta = 0.1, T_1(0)/T_2 = 0.01$; thick dashed line: $\beta = 0.999, T_1(0)/T_2 = 0.01$. $T_1(0)$ corresponds to the moment $t = 0$.

Fig. 4

Temporal dependence of $I_1/I_2$ according to (44) with allowance for temporal dependences of $T_1/T_2$ in Fig. 3. The correspondence of thick and solid lines is the same as in Fig. 3. The $\tau_{ST}$ corresponds to the moment of reaching thermal equilibrium (the saturation points on the curves shown in Fig. 3).